\documentclass[nofootinbib,floatfix,superscriptaddress,prd,twocolumn]{revtex4-1}
\usepackage{graphicx}
\usepackage{epsfig}
\usepackage{bm}
\usepackage{comment}
\pdfoutput=1
\usepackage[T1]{fontenc}
\usepackage[latin9]{inputenc}
\usepackage{amssymb}
\usepackage{float}
\usepackage{amsmath}
\usepackage{dcolumn}
\usepackage{cancel}

\usepackage[dvipsnames]{xcolor}

\usepackage[colorlinks]{hyperref}
\hypersetup{
    breaklinks=true,
    pdfstartview={FitH},    
    colorlinks=true,       
    linkcolor=blue,          
    citecolor=red,        
    filecolor=magenta,      
    urlcolor=magenta,           
    anchorcolor=green,      
    linktocpage=true
}

\newcommand{\be}{\begin{equation}}
\newcommand{\ee}{\end{equation}}

\begin{document}

\title{Holographic dark energy from a new two-parameter entropic functional}

\author{G.~G.~Luciano}
\email{giuseppegaetano.luciano@udl.cat}
\affiliation{Department of Chemistry, Physics and Environmental and Soil 
Sciences, Escola Politecninca Superior, Universidad de Lleida, Av. Jaume
II, 69, 25001 Lleida, Spain}
\author{E. N. Saridakis}
\email{msaridak@noa.gr}
 \affiliation{Institute for Astronomy, Astrophysics, Space Applications and 
Remote Sensing, National Observatory of Athens, 15236 Penteli, Greece}
 \affiliation{Departamento de Matem\'{a}ticas, Universidad Cat\'{o}lica del 
Norte, 
Avda.
Angamos 0610, Casilla 1280 Antofagasta, Chile}

\affiliation{CAS Key Laboratory for Researches in Galaxies and Cosmology, 
Department of Astronomy, University of Science and Technology of China, Hefei, 
Anhui 230026, P.R. China}

\begin{abstract}
We formulate an extended holographic dark energy scenario based on a recently 
proposed two-parameter generalized entropic functional. Unlike constructions 
that phenomenologically impose modified entropy-area relations at the horizon 
level, the present framework is rooted in a microscopic entropy functional and 
the corresponding microstate counting. For bounded systems, the entropy 
acquires 
a generalized holographic scaling with two independent area contributions, 
recovering the Bekenstein-Hawking entropy in the appropriate limits. 
Implementing this entropy within the holographic principle, we derive a 
generalized dark energy density containing two distinct holographic sectors, 
naturally embedding standard holographic dark energy and $\Lambda$CDM as 
limiting cases. We analyze the cosmological evolution for both Hubble and 
future 
event horizon cutoffs and show that the model successfully reproduces the 
matter-to-dark-energy transition. The two entropic exponents enrich the 
dynamics, allowing for quintessence-like behavior or phantom regimes, while 
remaining compatible with the standard thermal history of the Universe.
\end{abstract}

 \maketitle

\section{Introduction}
 
It is by now firmly established that the late-time Universe has undergone a 
transition from a matter-dominated phase to a period of accelerated expansion. 
While the simplest and most economical explanation is the presence of a 
cosmological constant, this interpretation is accompanied by well-known 
conceptual difficulties. In particular, the severe discrepancy between the 
observed value of the cosmological constant and the estimate from quantum field 
theory, as well as the possibility that the accelerated expansion may be 
dynamical rather than strictly constant, has motivated the exploration of 
alternative theoretical frameworks. 

Broadly speaking, two main directions have been pursued. The first retains 
general relativity as the fundamental theory of gravity and attributes cosmic 
acceleration to new, exotic components in the matter sector, collectively 
described as dark energy \cite{Copeland:2006wr,Cai:2009zp,Bamba:2012cp}. The 
second modifies or extends the gravitational sector itself, constructing 
theories that reduce to general relativity at low energies but introduce 
additional degrees of freedom capable of driving late-time acceleration 
\cite{Nojiri:2010wj,Capozziello:2011et,Cai:2015emx,CANTATA:2021asi}.

An alternative and conceptually distinct approach is provided by holographic 
dark energy. This scenario originates from the holographic principle 
\cite{tHooft:1993dmi,Susskind:1994vu,Bousso:2002ju}, applied in a 
cosmological context \cite{Fischler:1998st,Bak:1999hd,Horava:2000tb}. The 
central idea is that the validity of a quantum field theory at large 
distances requires an infrared cutoff, which is related to its ultraviolet 
cutoff through gravitational constraints, and ultimately to the vacuum energy 
itself \cite{Cohen:1998zx,Addazi:2021xuf}. 

In this framework, the vacuum energy 
acquires a holographic origin, naturally giving rise to a dark energy component 
whose density is determined by a cosmological length scale 
\cite{Li:2004rb,Wang:2016och}. Holographic dark energy has been extensively 
studied and shown to exhibit a rich and viable cosmological 
phenomenology, consistent with the thermal history of the Universe and current 
observational constraints 
\cite{Li:2004rb,Wang:2016och,Huang:2004ai,Pavon:2005yx,
Wang:2005jx,Nojiri:2005pu,Kim:2005at,Setare:2006wh,Setare:2008hm,Sheykhi:2009dz,
Li:2009bn,
Zhang:2009un,Micheletti:2010cm,Duran:2010hi,Lu:2009iv,Micheletti:2009jy,
Aviles:2011sfa,Zhai:2011pp,
Cardenas:2010wx,Chimento:2011pk,Zhang:2012uu,Cruz:2011wx,Huang:2012xma,
Zhang:2015rha,Nastase:2016sji,Mukherjee:2016lor,Zhao:2017urm,
Luongo:2017yta,Mukherjee:2017oom,
Pinki:2022aht,Mohammadi:2022vru,Landim:2022jgr,Trivedi:2024inb,Maity:2024tkq,
Li:2024bwr}.

Extensions of the basic holographic dark energy scenario typically follow two 
main directions. The first concerns the choice of the infrared cutoff, namely 
the cosmological length scale entering the holographic construction. Various 
possibilities have been explored in the literature, including the future event 
horizon, the apparent horizon, the age of the Universe, the conformal time, as 
well as curvature-based scales associated with the Ricci scalar or 
the Gauss-Bonnet invariant 
\cite{Gong:2004fq,Saridakis:2007cy,Cai:2007us,Setare:2008bb,Gong:2009dc,
Suwa:2009gm,Jamil:2010vr,Bouhmadi-Lopez:2011qvd,
Malekjani:2012bw,Landim:2015hqa,Rao:2017lnt,Shekh:2021ule,Rudra:2022qbv}. The 
second direction focuses on modifying the entropy-area relation of the 
cosmological horizon itself. Motivated by generalized statistical mechanics and 
quantum-gravitational considerations, a variety of extended entropy functionals 
have been proposed, including Tsallis, Barrow, Kaniadakis, logarithmically 
corrected, and other generalized entropies, leading to corresponding extensions 
of holographic dark energy 
\cite{Pasqua:2015bfz,Jawad:2016tne,Pourhassan:2017cba,Saridakis:2017rdo,
Nojiri:2017opc,Saridakis:2018unr,DAgostino:2019wko,Saridakis:2020zol,
Dabrowski:2020atl,daSilva:2020bdc,Anagnostopoulos:2020ctz,Mamon:2020spa,
Bhattacharjee:2020ixg,Huang:2021zgj,Drepanou:2021jiv,Hossienkhani:2021emv,
Nojiri:2021iko,Jusufi:2021fek,Hernandez-Almada:2021aiw,Hernandez-Almada:2021rjs,
Luciano:2022pzg,Yarahmadi:2024oqv,Cimdiker:2025vfn}.

In light of these developments, a novel generalized entropy functional that 
departs from the standard holographic scaling by incorporating two independent 
entropic exponents was recently introduced in \cite{Luciano:2026ufu}. Unlike 
approaches that impose modified entropy-area 
relations phenomenologically at the horizon level, this construction is rooted 
in a well-defined microscopic framework based on a generalized entropic 
functional and the corresponding microstate counting. 

When applied to systems 
with boundaries, the resulting entropy exhibits a generalized holographic-type 
behavior, leading to an extended area dependence that encompasses the 
Bekenstein-Hawking entropy as a particular limiting case. Furthermore, within 
the context of the gravity-thermodynamics correspondence, the use of this 
entropy modifies the cosmological evolution equations, giving rise to 
additional 
contributions that can be consistently interpreted as an effective dark energy 
sector of entropic origin \cite{Luciano:2026ufu, Leizerovich:2026pfy}.

In the present work, we formulate a holographic dark energy scenario based on 
the aforementioned two-parameter generalized entropy. By implementing the 
holographic principle in a cosmological framework and employing the extended 
entropy-area relation, we derive the corresponding dark energy density and 
construct the associated cosmological equations. We investigate the resulting 
background dynamics in a homogeneous and isotropic Universe, focusing on the 
evolution of the dark energy density parameter and the dark energy 
equation-of-state parameter. Particular attention is devoted to the limiting 
cases, in which the scenario consistently reproduces standard holographic dark 
energy as well as the $\Lambda$CDM paradigm. Our analysis shows that the 
generalized entropic structure leads to a richer cosmological behavior while 
remaining compatible with the standard thermal history of the Universe.

The plan of the paper is as follows. In Section~\ref{themodel} we briefly 
review 
the new two-parameter entropic functional and discuss its generalized 
holographic scaling, emphasizing its microscopic origin and the limiting cases 
that recover the Bekenstein-Hawking entropy. We then implement this entropy 
within the holographic dark energy framework and derive the corresponding 
generalized dark energy density. In Section~\ref{evolution} we investigate the 
cosmological evolution of the resulting scenario, examining both the Hubble 
horizon and the future event horizon as infrared cutoffs, and analyze the 
evolution of the dark energy density parameter and the dark energy 
equation-of-state parameter. Finally, in Section~\ref{Conclusions} we summarize 
our results and discuss possible future extensions of the present framework.

\section{Holographic dark energy from a new two-parameter entropic functional }
\label{themodel}

In this section  we construct an extended holographic dark energy scenario 
based on the recently proposed two-parameter generalized entropic functional. 
Our goal is to consistently implement the modified entropy within the 
holographic framework and derive the corresponding dark energy density 
and cosmological equations. 

We begin by briefly reviewing the entropic functional and its associated 
generalized microstate scaling, emphasizing its microscopic origin and 
holographic-type behavior for systems with boundaries. Subsequently, we 
incorporate this entropy into the holographic principle, replace the standard 
Bekenstein-Hawking area law in the holographic bound, and obtain the resulting 
generalized holographic dark energy density. Finally, we derive the 
corresponding Friedmann equations and discuss the relevant limiting cases in 
which standard holographic dark energy and $\Lambda$CDM are recovered.

\subsection{A new two-parameter entropic functional}

 In standard statistical mechanics, equilibrium systems are described by the
Boltzmann-Gibbs-Shannon (BGS)  entropy
$
S_{\text{BGS}}=-\kappa\sum_{i=1}^{W}p_i\ln p_i$,
with $p_i$   the probability of the $i$-th microstate among the $W$
accessible ones, and $\kappa$  the entropy units.  Nevertheless, one can write 
a more general class of entropic measures   as 
\cite{hanel2011comprehensive}
\begin{equation}
\label{Sf}
S_f[p]=\sum_i f(p_i)\,.
\end{equation}
Imposing continuity, maximality, expandability, and separability, i.e. the four 
Khinchin axioms, uniquely selects the Boltzmann-Gibbs-Shannon entropy 
\cite{shannon1948claude,Khinchin1957}. However, relaxing these axioms leads to 
generalized entropic measures.

In particular, if $L$ is a characteristic length scale of the system, then the 
Boltzmann-Gibbs-Shannon entropy obeys the holographic-type scaling 
\cite{tHooft:1993dmi,Susskind:1994vu}
$S_{\text{BGS}}\propto \log W \propto L^2$, which implies a microstate 
counting of the form
$W = g(L)\,\xi^{L^2}$ with $\xi>1$, 
where $g(L)$ is subleading at large $L$.

In \cite{Luciano:2026ufu} we introduced the generalized entropy functional as
\begin{eqnarray}
\nonumber
S_{\delta,\epsilon}&=&\eta_\delta \sum_i 
p_i\left(\log\frac{1}{p_i}\right)^\delta
+\eta_\epsilon \sum_i p_i\left(\log\frac{1}{p_i}\right)^\epsilon\\[1mm]
&=&\eta_\delta\left(\log W\right)^\delta
+\eta_\epsilon\left(\log W\right)^\epsilon\,,
\label{genentropyexpr}
\end{eqnarray}
for equiprobable distributions, where $\eta_\delta$ and
$\eta_\epsilon$ 
are parameters, with $\delta,\epsilon>0$,   
corresponding to the generalized microstate scaling
\begin{equation}
\label{microscaling}
W_{\delta,\epsilon}
= g(L)\,\xi^{L^{2\delta}}\,\tilde{\xi}^{L^{2\epsilon}}\,,
\qquad \xi,\tilde{\xi}>1\,.
\end{equation}
Hence, for systems with a boundary of area $A\sim L^2$, it leads at large 
$L$ to the holographic-like entropy \cite{Luciano:2026ufu}
\begin{equation}
\label{Sde}
S_{\delta,\epsilon}
=\gamma_\delta A^\delta+\gamma_\epsilon A^\epsilon\,,
\end{equation}
where $\gamma_\delta$ and $\gamma_\epsilon$ are positive constants with
dimensions $[L^{-2\delta}]$ and $[L^{-2\epsilon}]$, respectively. The pair
$(\delta,\epsilon)$ therefore parametrizes deviations from the standard
holographic scaling through two independent terms.

We stress that the above modified entropy arises from a modification of the 
microscopic functional and the associated
microstate scaling, rather than being introduced directly at the macroscopic 
level. Moreover, the Bekenstein-Hawking entropy is 
recovered in three cases, namely: \emph{i)} $\delta=1,\ 
\gamma_\delta=\frac{1}{4\ell_p^2},\ \gamma_\epsilon=0$, \emph{ii)} 
$\epsilon=1,\ 
\gamma_\delta=0,\ \gamma_\epsilon=\frac{1}{4\ell_p^2}$, and \emph{iii)} 
$\delta=\epsilon=1,\ 
\gamma_\delta=\gamma_\epsilon=\frac{1}{8\ell_p^2}$.

\subsection{Holographic dark energy }
\label{HDEl}

Let us now employ the above generalized entropy to construct an extended 
holographic dark energy scenario. Throughout this work, we consider a flat, 
homogeneous and isotropic Friedmann-Robertson-Walker (FRW) metric of the form
\begin{equation}
\label{FRc}
ds^{2}=-dt^{2}+a^{2}(t)\delta_{ij}dx^{i}dx^{j}\,,
\end{equation}
where $a(t)$ is the scale factor.

In the framework of holographic dark energy, one imposes the
inequality $\rho_{DE} L^4 \leq S$ with $S\propto A \propto L^2$, where 
$\rho_{DE}$ is the dark energy 
density of holographic origin, $L$ is a horizon of the Universe, and $A$ its 
area \cite{Wang:2016och}. Hence, the Friedmann equation reads
\begin{eqnarray}
\label{Fr1bFRW}
3M_p^2 H^2 = \rho_m + \rho_{DE},
\end{eqnarray}
where $\rho_{DE}=3c^2 M_p^2 L^{-2}$ (here $M_p$ 
is the Planck mass), $c$ is a constant, $\rho_m$ is the matter energy 
density, 
and $H\equiv \dot{a}/a$ is the Hubble parameter.
  
We follow the same procedure, but instead of the standard 
Bekenstein-Hawking entropy we employ the generalized entropy (\ref{Sde}). In 
this 
case, the Friedmann equation is again given by Eq. (\ref{Fr1bFRW}), but we now 
obtain
\begin{eqnarray}
\label{GenHDE}
&&\rho_{DE}=\frac{ \gamma_\delta 
\left(L^2\right)^\delta\,+\,\gamma_\epsilon 
\left(L^2\right)^\epsilon}{L^4}\nonumber\\
&&=  
 \gamma_\delta 
L^{2(\delta-2)}
+\gamma_\epsilon 
L^{2(\epsilon-2)}.
\end{eqnarray}
Note that one could also introduce a multiplicity coefficient $c$. However, we 
have 
absorbed it into $\gamma_\delta$ and $\gamma_\epsilon$ (keeping the same 
symbols 
for simplicity). As one can see, in the three limiting cases where 
(\ref{Sde}) reduces to the standard Bekenstein-Hawking entropy, Eq. 
(\ref{GenHDE}) reproduces standard holographic dark energy 
\cite{Li:2004rb,Wang:2016och}. In the general case, however, we obtain an 
extended 
holographic dark energy scenario. Lastly, it is worth mentioning that in the 
special cases where $\gamma_\delta=0$, $\epsilon=2$, or $\gamma_\epsilon=0$, 
$\delta=2$, or $\delta=\epsilon=2$, the above relation yields the standard 
cosmological constant case $\rho_{DE}=\mathrm{const.}=\Lambda$.

In order to proceed, one needs to consider an explicit infrared cutoff, namely 
a specific horizon scale $L$. In the basic holographic dark energy models, $L$ 
cannot 
be the Hubble horizon $H^{-1}$, since such a choice leads 
to inconsistencies \cite{Hsu:2004ri}. For this reason, one typically employs 
the 
future 
event horizon \cite{Li:2004rb}, i.e.
\begin{equation}
\label{futurehor}
R_h\equiv a(t)\int_t^\infty \frac{dt'}{a(t')}= a\int_a^\infty 
\frac{da'}{H(a')\hspace{0.2mm}a'^2}.
\end{equation}
However, in generalized holographic dark energy scenarios one may instead adopt 
other horizon scales as well, such as the Hubble horizon, the Ricci scalar, the 
Ricci-Gauss-Bonnet scalar, the Granda-Oliveros cutoff, etc. In the 
following, 
we present two applications, namely one with the Hubble horizon and one with 
the future event horizon.

\subsubsection{Holographic dark energy using the Hubble horizon}

If we consider the infrared cutoff to be the Hubble horizon $H^{-1}$, then  
Eq. (\ref{GenHDE}) becomes
 \begin{equation}
\label{GenHDEHubble}
\rho_{DE}=  
 \gamma_\delta 
H^{2(2-\delta)}
+\gamma_\epsilon 
H^{2(2-\epsilon)}\,.
\end{equation}
Now, since the matter sector is conserved, the dark energy sector is 
conserved as well, i.e. we have
\begin{equation}
\label{rhodeconserv}
\dot{\rho}_{DE}+3H\rho_{DE}(1+w_{DE})=0,
\end{equation}
where $w_{DE}$ is the holographic dark energy equation-of-state parameter.
  
Hence, differentiating 
Eq. (\ref{GenHDEHubble}) and inserting into Eq. (\ref{rhodeconserv}), 
we obtain
\begin{equation}\label{wderelation}
 w_{DE}=-1+\frac{2 \dot{H}[\gamma_\epsilon (\epsilon-2 ) H^{2\delta}+
 \gamma_\delta (\delta-2 ) H^{2\epsilon}]}
 {3H^2(\gamma_\epsilon  H^{2\delta} + \gamma_\delta H^{2\epsilon} )}.
\end{equation}
In the special case where $\epsilon=2$, we obtain
\begin{eqnarray}
\label{GenHDEHubblespec1}
&&\rho_{DE}=  
 \gamma_\delta 
H^{2(2-\delta)}
+\gamma_\epsilon \,,
\end{eqnarray}
which corresponds to a correction to the $\Lambda$CDM scenario, while taking 
$\gamma_\delta=0$ 
recovers the $\Lambda$CDM scenario exactly. Similarly, in the special 
case $\delta=2$, we obtain
\begin{eqnarray}
\label{GenHDEHubblespec2}
&&\rho_{DE}=  
 \gamma_\epsilon
H^{2(2-\epsilon)}
+\gamma_\delta \,, 
\end{eqnarray}
which again corresponds to a correction to the $\Lambda$CDM scenario, 
recovering 
the $\Lambda$CDM 
paradigm for $\gamma_\epsilon=0$.

\subsubsection{Holographic dark energy using the  future event horizon }
If we consider the infrared cutoff to be the future event horizon 
(\ref{futurehor}), then 
Eq. (\ref{GenHDE}) gives
\begin{eqnarray}
\label{GenHDE2}
&&\rho_{DE}= 
 \gamma_\delta 
R_h^{2(\delta-2)}
+\gamma_\epsilon 
R_h^{2(\epsilon-2)}.
\end{eqnarray}
It is convenient to introduce the dark matter and dark energy density 
parameters as
\begin{eqnarray}
 && \Omega_m\equiv\frac{1}{3M_p^2H^2}\rho_m
 \label{Omm}\\
 &&\Omega_{DE}\equiv\frac{1}{3M_p^2H^2}\rho_{DE}.
 \label{ODE}
\end{eqnarray}

Hence, inserting Eq.~(\ref{futurehor}) into (\ref{GenHDE2}) and then into the 
Friedmann equation (\ref{Fr1bFRW}) gives
\begin{eqnarray}
 && \!\!\!\!\!\!\!\!\!\!\! 
\frac{3M_p^2H^2}{\gamma_\delta}e^{-2(\delta-2)x}\Omega_{DE}=\left[\int_x^\infty 
\frac{dx}{Ha}\right]^{2(\delta-2)}\nonumber\\
&&
\ \ \ \ \ \ \ \ \ \ \  
+\frac{\gamma_\epsilon}{\gamma_\delta} 
e^{2(\epsilon-\delta)x}\left[\int_x^\infty 
\frac{dx}{Ha}\right]^{2(\epsilon-2)},\label{fullequa}
\end{eqnarray}
where $x\equiv \ln a$. 

The presence of the two different powers in the integrals prevents an 
analytical treatment. Hence, in the following we focus on the case 
$\epsilon=2$ (equivalently, one could consider $\delta=2$).
In this case, Eq. (\ref{GenHDE}) becomes
 \begin{equation}
\label{GenHDEepsilon2}
\rho_{DE}= 
 \gamma_\delta 
R_h^{2(\delta-2)}
+\gamma_\epsilon ,
\end{equation}
and thus Eq. (\ref{fullequa}) simplifies to  
 \begin{equation}
\label{HDERh22}
\frac{3M_p^2H^2}{\gamma_\delta}e^{-2(\delta-2)x}\Omega_{DE}=\left[\int_x^\infty 
\frac{dx}{Ha}\right]^{2(\delta-2)} +\frac{\gamma_\epsilon}{\gamma_\delta} 
e^{2(2-\delta)x}.
 \end{equation}
   
As usual, we consider the matter component to be dust, in which case the matter 
conservation equation gives 
$\rho_m=\rho_{m0}/a^3$, with $\rho_{m0}$ the current value of the matter 
energy density (in the following, the subscript ``0'' denotes the  
value of a quantity at present). 
Inserting this into Eq. (\ref{Omm}) yields 
$\Omega_m=\Omega_{m0} H_0^2/(a^3 H^2)$, which, using the Friedmann 
equation, leads to
\begin{equation}\label{Hrelhelp}
\frac{1}{Ha}=\frac{\sqrt{a(1-\Omega_{DE})}}{H_0\sqrt{\Omega_{m0}}}.
\end{equation}

On the other hand, one has
  \begin{equation}
  \label{Hrelhelp2}
 \frac{d}{dx}\left[\int_x^\infty 
\frac{dx}{Ha}\right]=-\frac{1}{aH}.
 \end{equation}
Thus, differentiating Eq. (\ref{HDERh22}) with respect to $x$ and using 
the relations (\ref{Hrelhelp}) and (\ref{Hrelhelp2}), we finally acquire
\begin{widetext}
    \begin{eqnarray}\label{Odediffeqfull}
&&
\!\!\!\!\!\!\!\!\!\!\!\!\!\!\!\! 
3M_p^2 h_0^3 \frac{\Omega_{DE}'}{ (\Omega_{DE}-1)}
= \Omega_{DE} \left\{3 (1-2 
\delta ) M_p^2 h_0 ^3+6 (\delta -2) M_p^2 h_0 ^2  
\sqrt{1-\Omega_{DE}} \left[\frac{3 M_p^2 h_0 ^2
   e^{3(1-\delta ) x}\Omega_{DE}}{\gamma_\delta (1-  
\Omega_{DE})}-\frac{\gamma_\epsilon}{\gamma_\delta} 
   e^{3(2-\delta ) x}\right]^{\frac{1}{4-2 \delta }}\right.\nonumber\\
&&
\ \ \ \ \ \ \ \ \ \ \ \ \ \ \ \ \ \ \ \ \ \ \  
\left.
+2 \gamma_\epsilon  (\delta
   -2) e^{7 x/2} \sqrt{1-\Omega_{DE}} \left[\frac{3 M_p^2 h_0 ^2 e^{x-2 
\delta  x} \Omega_{DE}}{\gamma_\delta (1-  
\Omega_{DE})}-\frac{\gamma_\epsilon}{\gamma_\delta}   e^{-2
   (\delta -2) x}\right]^{\frac{1}{4-2 \delta }}-2\gamma_\epsilon    (\delta 
-2) 
h_0  e^{3 x}\right\}\nonumber\\ 
&& \ \ \ \ \ \ \ \ \ \ \ \ \ \ \ \ \ \  \
-2 \gamma_\epsilon   (\delta -2) e^{3 x}
   \left\{e^{x/2} \sqrt{1-\Omega_{DE}} \left[\frac{3 M_p^2 h_0 ^2 e^{x-2 
\delta  x} \Omega_{DE}}{\gamma_\delta (1-  
\Omega_{DE})}-\frac{\gamma_\epsilon}{\gamma_\delta}   e^{-2
   (\delta -2) x}\right]^{\frac{1}{4-2 \delta }}-h_0 \right\},
\end{eqnarray}
\end{widetext}
where $h_0=H_0\sqrt{\Omega_{m0}}$.
 
Equation (\ref{Odediffeqfull}) is the differential equation that determines the 
evolution of  the generalized holographic dark energy density $\Omega_{DE}(x)$, 
in a flat universe 
and for dust matter, as a function of $x$. 
 We mention here that for $\gamma_\epsilon=0$ the above expression simplifies 
to 
  \begin{eqnarray}\label{Odediffeq}
&&
\!\!\!\!\!\!\!\!\!\!\!
\frac{\Omega_{DE}'}{\Omega_{DE}(1-\Omega_{DE})}=2\delta-1+
Q
(1-\Omega_{DE})^{\frac{1-\delta}{
2(2-\delta) } } \nonumber\\
&&\ \ \ \ \ \ \ \ \ \ \ \ \ \  \ \ \ \ \ \ \ \ \ \  \  \ \ \   \ \ \ 
\cdot(\Omega_{DE})^{\frac{1}{2(2-\delta) } } 
e^{\frac{3(1-\delta)}{2(2-\delta)}x},
\end{eqnarray}
 with
   \begin{equation}\label{Qdef}
Q\equiv 
2(2-\delta)\left(\frac{\gamma_\delta}{3M_p^2}\right)^{\frac{1}{2(\delta-2)}} 
\left(H_0\sqrt{\Omega_{m0}}\right)^{\frac{1-\delta}{\delta-2}},
\end{equation}
namely it coincides with Tsallis \cite{Saridakis:2018unr} and Barrow 
\cite{Saridakis:2020zol}  holographic dark energy.
Finally, in the case $\gamma_\epsilon=0$ and $\delta=1$, 
the model coincides with standard holographic dark energy \cite{Li:2004rb}, 
namely
$\Omega_{DE}' = 
\Omega_{DE}(1-\Omega_{DE})\left(1+2\sqrt{\frac{3M_p^2\Omega_{DE}}{\gamma_\delta}
}
\right)$
(complete agreement is obtained upon the identification 
$\gamma_\delta=3 c^2 M_p^2$).

We close this subsection by deriving an analytical expression for the 
holographic dark energy equation-of-state parameter $w_{DE}$. As mentioned, 
since the 
matter sector is conserved, the dark energy sector is conserved as well.
Now, differentiating   Eq. (\ref{GenHDEepsilon2}) we obtain   
\begin{equation}
\dot{\rho}_{DE}=2(\delta-2)\gamma_\delta  R_h^{2\delta-5} \dot{R}_h,
\end{equation}
and 
$\dot{R}_h$ can be easily
  found from Eq. (\ref{futurehor}) to be 
$\dot{R}_h=H  R_h-1$,  while $R_h$ can be eliminated in terms of 
$\rho_{DE}$ 
through 
\begin{equation}  
R_h=\left(\frac{\rho_{DE}-\gamma_\epsilon}{\gamma_\delta}\right)^{\frac{1}{ 
2(\delta-2)}} \,,
\end{equation}
 which arises from  Eq. (\ref{GenHDEepsilon2}).
 
Hence, inserting the above expressions into Eq. (\ref{rhodeconserv}),   we find 
\begin{eqnarray}\label{rhodeconserv2}
&&
\!\!\!\! 
2(\delta\! -\! 2)\gamma_\delta 
\left(\frac{\rho_{DE}\! -\! \gamma_\epsilon}{\gamma_\delta 
}\right)^{\frac{2\delta-5}{2(\delta-2)}}
 \left[H  
\left(\frac{\rho_{DE}\! -\! \gamma_\epsilon}{\gamma_\delta 
}\right)^{\frac{1}{2(\delta-2)}}-1\right]\nonumber\\
&&\ \ \ \ \ \  \ \  \ \ \ \ \  \ \ \ \ \ \ \ \ \  \ \  \ \ \ \ \  
+3H\rho_{DE}
(1+w_{DE})=0.
\end{eqnarray}
If we substitute $H$ from Eq. (\ref{Hrelhelp}) and use relation (\ref{ODE}), 
we 
finally extract  
\begin{widetext}
\begin{eqnarray}\label{wDE}
 &&
 \!\!\!\! \!\!\!\! \!\!\!\! \!\!\!\! \!\!\!\! \!\!\!\! 
 w_{DE}= -1+2(2-\delta)\gamma_\delta \left[\frac{3 M_p^2 h_0^2\Omega_{DE} 
e^{-3x}-\gamma_\epsilon (1-\Omega_{DE})}{\gamma_\delta(1-\Omega_{DE})}
\right]^{\frac{2\delta-5}{2(\delta-2)}}\cdot\frac{(1-\Omega_{DE})e^{3x}}{9M_p^2 
h_0^2\Omega_{DE}}\nonumber\\
 && \ \ \ \ \ \ \ \ \ \ \ \ \ \ \ \ \ \ 
 \cdot \left[\left(\frac{3 M_p^2 h_0^2\Omega_{DE} 
e^{-3x}-\gamma_\epsilon 
(1-\Omega_{DE})}{\gamma_\delta(1-\Omega_{DE})}\right)^{ {\frac{1}
{2(\delta-2)}}}
-\frac{(1-\Omega_{DE})^{ {1/2}}e^{3x/2}}{h_
0}
\right].
\end{eqnarray}

\end{widetext}
 Once again, if  
 $\gamma_\epsilon=0$ the above expression simplifies 
to 
\begin{equation}\label{wDEbis}
w_{DE}=\frac{1-2\delta}{3}
-\frac{Q}{3}
(\Omega_{DE})^{\frac{1}{2(2-\delta) } } (1\!-\!\Omega_{DE})^{\frac{\delta-1}{
2(\delta-2) } }
e^{\frac{3(1-\delta)}{2(2-\delta)}x}.
\end{equation}
which is just the result of   Tsallis \cite{Saridakis:2018unr} and Barrow 
\cite{Saridakis:2020zol}   holographic dark 
energy.
Lastly, in the case $\gamma_\epsilon=0$ and $\delta=1$
it recovers the expression of standard holographic dark energy  
\cite{Li:2004rb},  
namely 
$w_{DE}|_{\delta=1}=-\frac{1}{3}-\frac{2}{3}\sqrt{\frac{3M_p^2 
\Omega_{DE}}{\gamma_\delta}}$ 
\cite{Wang:2016och}, where complete coincidence is acquired under the 
identification  
$\gamma_\delta=3 
c^2 M_p^2$.

\section{Cosmological evolution}
\label{evolution}

In this section, we study in detail the cosmological evolution of the 
extended holographic dark energy scenario that arises from the new 
two-parameter entropic functional. In particular, we examine both the case 
where the IR cutoff is the Hubble horizon and the case where it is the 
future event horizon. Since analytical solutions are not available in the 
general case, we proceed numerically. As usual, we use the relation 
$a/a_0=(1+z)^{-1}$, where $z$ is the redshift, and we set $a_0=1$ at 
present. Moreover, concerning the initial conditions, we impose 
$\Omega_{DE}(x=0)\equiv\Omega_{DE0}\approx0.7$ and thus 
$\Omega_m(x=0)\equiv\Omega_{m0}\approx0.3$, in agreement with current 
observations.

\subsection{Hubble horizon case}

 \begin{figure}[h]
\includegraphics[scale=0.33]{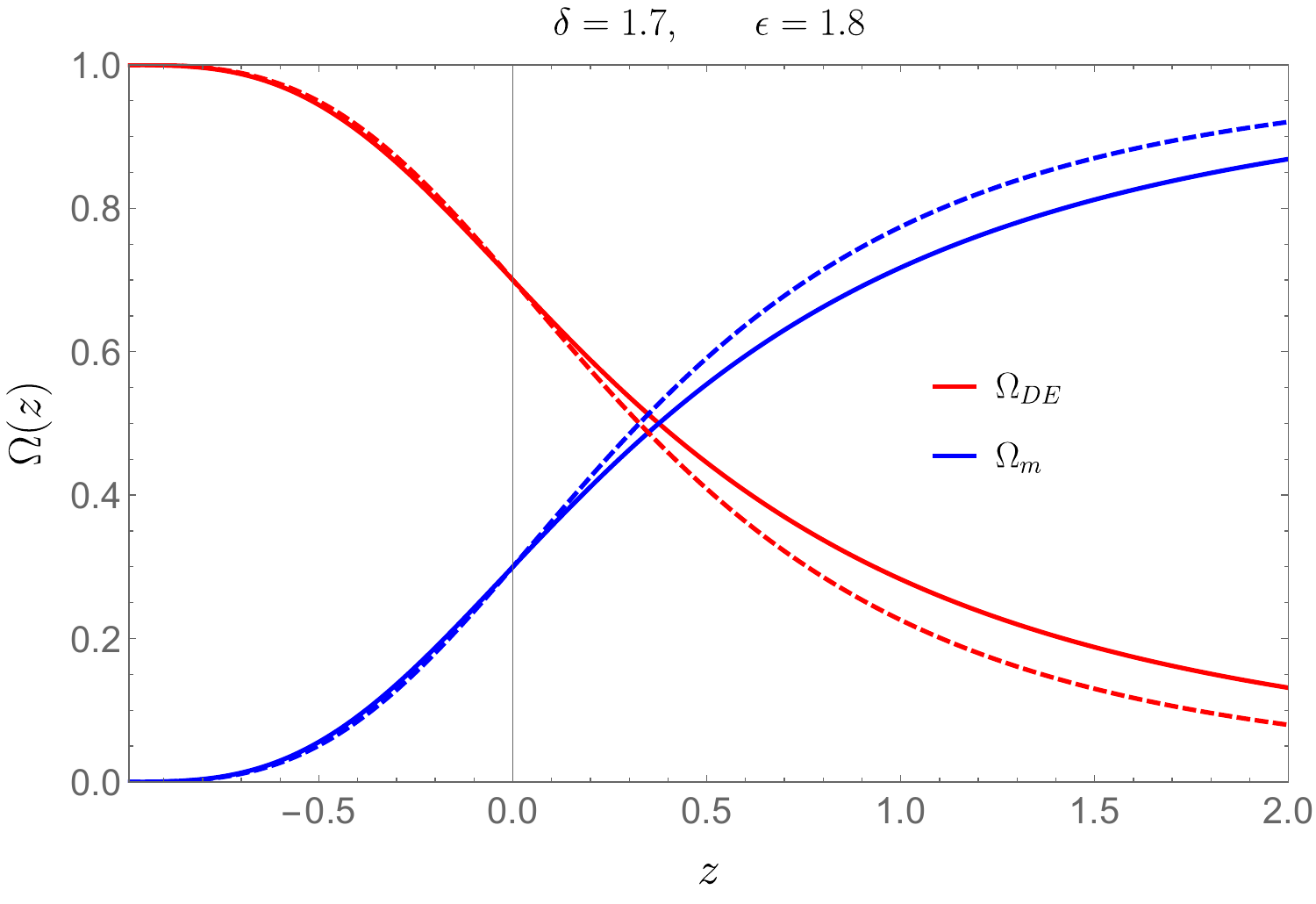}\\
\includegraphics[scale=0.33]{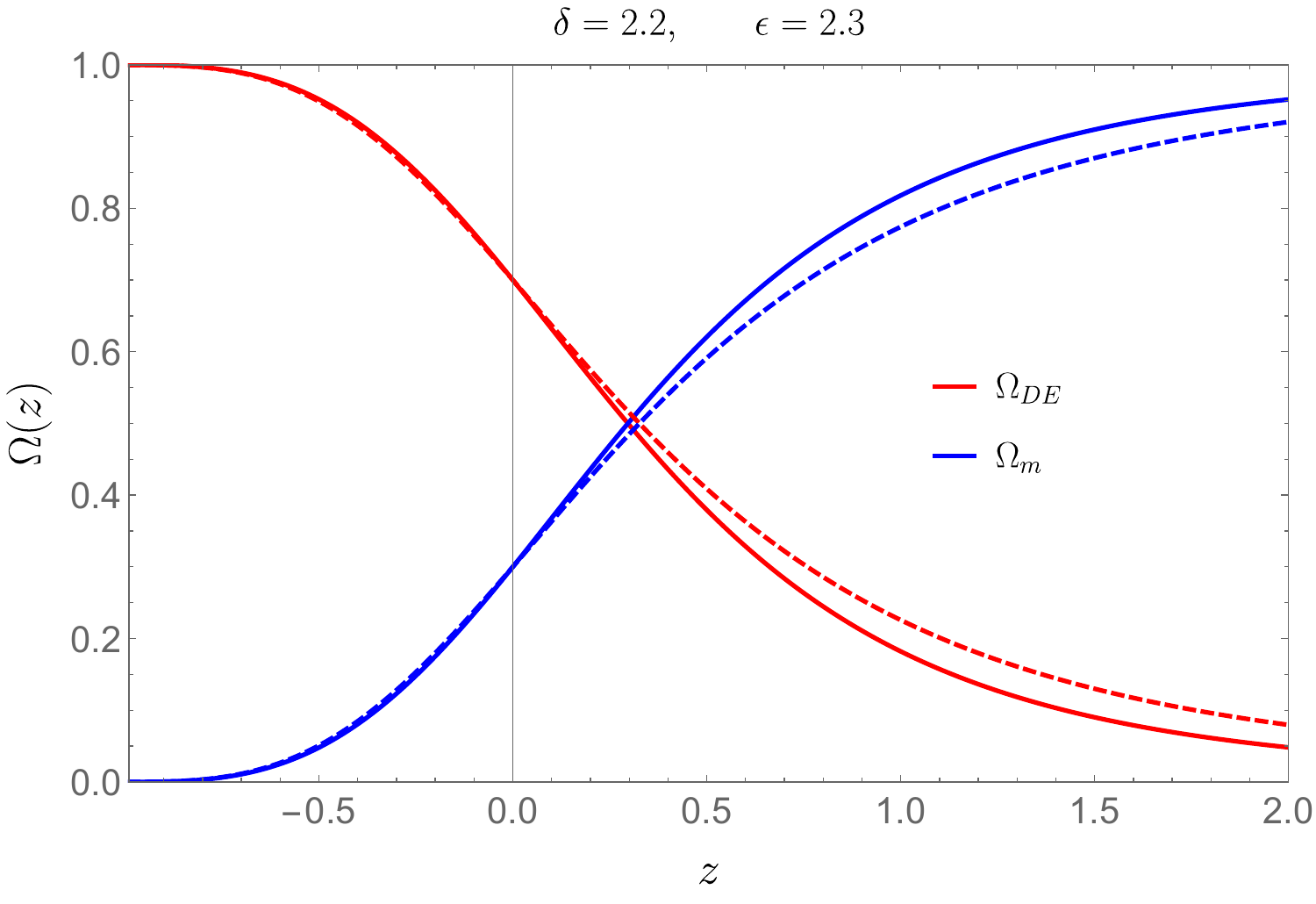}
\caption{
{\it{ Evolution of the matter and holographic dark energy density parameters as 
functions of the redshift $z$, assuming the Hubble horizon as the IR cutoff.
\textbf{Upper panel:} holographic dark energy with $\delta=1.7$, 
$\epsilon=1.8$, 
$\gamma_\delta=1$ and $\gamma_\epsilon$ determined by requiring 
$\Omega_{DE}(z=0)\equiv\Omega_{DE0}\approx0.7$ at the present epoch (solid 
curve), compared with the $\Lambda$CDM evolution (dashed curve).
\textbf{Lower panel:} same setup with $\delta=2.2$ and $\epsilon=2.3$. Units 
are 
chosen such that $M_p^2=1$.}} }
\label{OmegaFRWs}
\end{figure}

We first examine the scenario in which the infrared cutoff is chosen to be the 
Hubble horizon $L=H^{-1}$. In this case, the dark energy density is given by 
Eq.~(\ref{GenHDEHubble}), and the equation-of-state parameter follows from 
Eq.~(\ref{wderelation}). 

In Fig.~\ref{OmegaFRWs} we present the behavior of $\Omega_{DE}(z)$ for 
various choices of $(\delta,\epsilon)$, together with the $\Lambda$CDM 
evolution. 
We observe that the model can reproduce the standard matter-to-dark-energy 
transition. 
In contrast to standard holographic dark energy, where the Hubble horizon 
choice 
leads to 
inconsistencies, the present generalized entropic structure allows for a viable 
cosmological 
evolution due to the modified scaling of $\rho_{DE}$.

The presence of two independent exponents allows for a controlled deformation 
of 
the standard evolution. In all cases the model closely reproduces the 
$\Lambda$CDM behavior around the present epoch ($z \lesssim 0$), where the 
curves nearly overlap. Differences become visible at positive redshift. For 
$\delta,\epsilon<2$ (upper panel), $\Omega_{DE}$ lies above the $\Lambda$CDM 
prediction at intermediate and high redshift, implying that dark energy was 
relatively more important in the past and therefore grows more mildly toward 
the 
future. Conversely, for $\delta,\epsilon>2$ (lower panel), $\Omega_{DE}$ 
remains 
below the standard curve at $z>0$, indicating a smaller past contribution and 
thus a steeper future increase toward dark-energy domination. In the limit 
$\delta,\epsilon\to2$, the evolution smoothly converges to the cosmological 
constant case.

\begin{figure}[t]
  \hspace{-1.cm}
\includegraphics[scale=0.36]{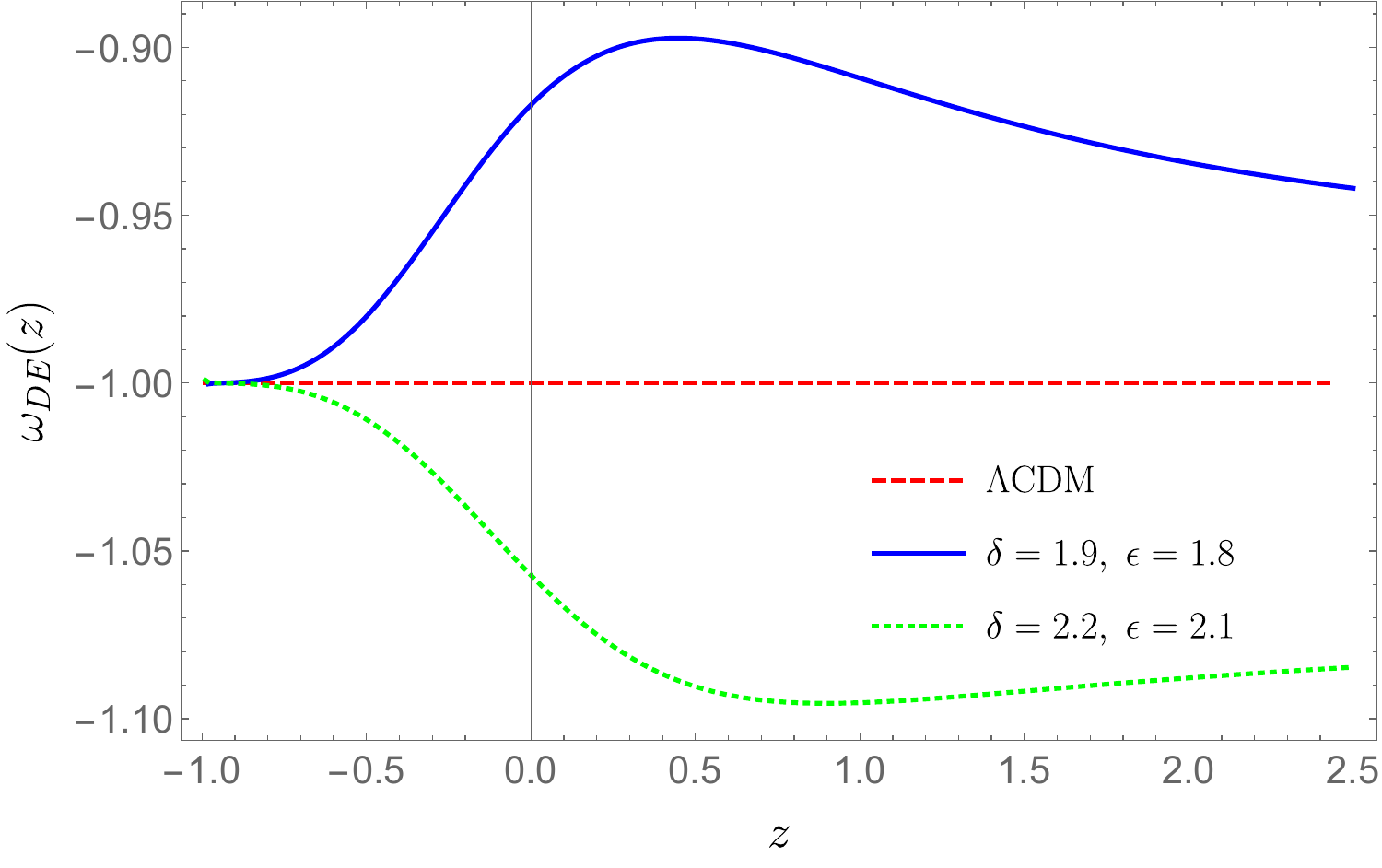}
\caption{
{\it{Evolution of the holographic dark energy equation-of-state parameter 
$w_{DE}$ as a function of the redshift $z$, assuming the Hubble horizon as the 
IR cutoff, for $\delta=1.9$, $\epsilon=1.8$ (solid blue curve) and 
$\delta=2.2$, 
$\epsilon=2.1$ (dotted green curve). In both cases, we have fixed 
$\gamma_\delta$ and $\gamma_\epsilon$ as in Fig.~\ref{OmegaFRWs}. The red 
dashed 
curve corresponds to the $\Lambda$CDM prediction. Units are chosen such that 
$M_p^2=1$.
}} }
\label{wdehub}
\end{figure}

Furthermore, in Fig.~\ref{wdehub} we show the evolution of $w_{DE}(z)$. 
Depending on the choice of parameters, the model can remain entirely within 
either the quintessence ($w>-1$) or the phantom ($w<-1$) regime during the 
cosmic evolution. 

\subsection{Future event horizon case}

We now consider the case in which the infrared cutoff is chosen to be the 
future event horizon. In this model, the generalized entropy leads to a dark 
energy density parameter that satisfies the differential equation 
(\ref{Odediffeqfull}).
 
\begin{figure}[t]
\includegraphics[scale=0.33]{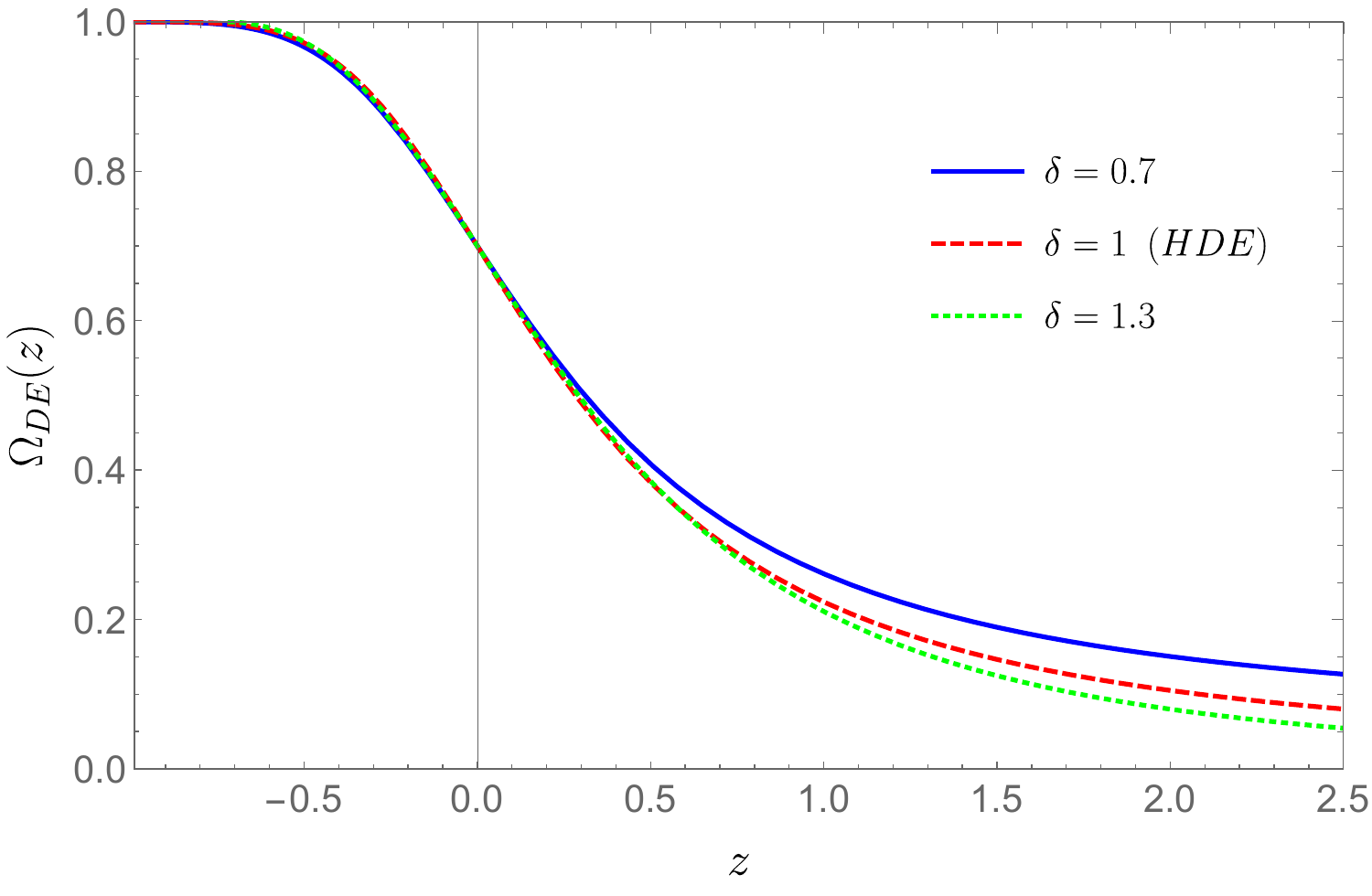}
\caption{
{\it{Evolution of the holographic dark energy density parameter as a function 
of 
the redshift $z$, assuming the future event horizon as the IR cutoff, for 
$\delta=0.7$ (solid blue curve) and $\delta=1.3$ (dotted green curve). In both 
cases, we have fixed $\epsilon=2$, $\gamma_\delta=0.05$, and have determined 
$\gamma_\epsilon$ by requiring $\Omega_{DE}(z=0)\equiv\Omega_{DE0}\approx0.7$ 
at 
the present epoch. The red dashed curve corresponds to the standard description 
of holographic dark energy. Units are chosen such that $M_p^2=1$.
}} }
\label{OmegaFRWsfuthoriz}
\end{figure}

In Fig.~\ref{OmegaFRWsfuthoriz} we present the evolution of $\Omega_{DE}(z)$ 
for representative values of the entropic parameters. As observed, the model 
successfully reproduces the standard thermal history of the Universe and the 
onset of dark-energy domination. The effect of the entropic exponents is to 
modify the rate at which the dark energy component 
overtakes matter. For values larger than unity (green curve), the onset of 
dark-energy domination occurs later, whereas for values smaller than unity 
(blue 
curve) 
dark energy is slightly enhanced. This behavior reflects the competing 
holographic scalings introduced by the two entropic terms, which effectively 
alter the redshift dependence of $\rho_{DE}$. 
Finally, we note that for $\delta=\epsilon=1$ we recover the standard 
holographic dark energy 
scenario, while for $\delta=\epsilon=2$ the $\Lambda$CDM paradigm is obtained.

Hence, the new two-parameter entropic functional leads to a richer cosmological 
behavior compared to single-exponent generalized holographic models, while 
remaining compatible with the observed thermal history of the Universe. 
In summary, the scenario successfully reproduces the sequence of cosmological 
epochs, and provides a broader framework in which $\Lambda$CDM and standard 
holographic dark energy emerge as limiting cases.

\section{Conclusions and Outlook}
\label{Conclusions}
 
In this work, we constructed an extended holographic dark energy scenario  
based on a recently proposed two-parameter entropic functional.  
Unlike approaches that phenomenologically impose modified entropy-area 
relations  
at the horizon level, this construction is rooted in a well-defined microscopic 
framework based on a generalized entropic functional  
and the corresponding microstate counting.  
The resulting entropy exhibits a generalized holographic-type behavior  
containing two independent area scalings and reduces to the  
Bekenstein-Hawking entropy in the appropriate limits.  
Implementing this entropy within the holographic dark energy  
construction, we derived a generalized dark energy density containing two  
distinct holographic contributions.  
In this way, the present framework provides a conceptually consistent bridge  
between microscopic entropy generalizations and late-time cosmological  
dynamics, while embedding standard holographic dark energy and $\Lambda$CDM as 
limiting  
cases.

At the background level, we analyzed the cosmological evolution  
for two representative choices of the infrared cutoff, namely the Hubble  
horizon and the future event horizon.  
In both cases, the model successfully reproduces the standard thermal history  
of the Universe, exhibiting a prolonged matter-dominated epoch followed by a  
late-time transition to accelerated expansion.  
Depending on the values of the entropic exponents $(\delta,\epsilon)$,  
the onset of dark-energy domination may occur earlier or later compared to  
the standard scenario, reflecting the interplay between the two holographic  
area scalings.  
Importantly, the construction remains fully consistent with the observed  
present-day values $\Omega_{DE0}\simeq0.7$ and $\Omega_{m0}\simeq0.3$,  
demonstrating its phenomenological viability.

Concerning the effective dark-energy equation-of-state parameter, we found that 
the  two-parameter entropic structure significantly enriches the cosmological  
behavior.  
For exponent values close to unity, the model behaves as a mild deformation of  
standard holographic dark energy, while for slightly larger values the  
effective equation-of-state parameter can enter the phantom regime during the 
evolution, without introducing additional  
pathological degrees of freedom.  
In the special cases where one of the exponents equals two, the dark energy  
density becomes constant and the scenario reduces exactly to $\Lambda$CDM.  
Moreover, when one of the entropic contributions vanishes, the model reproduces 
 Tsallis and Barrow holographic dark energy, and in the further limit  
$\delta=1$ the standard holographic dark energy case is recovered.  
Hence, the present construction constitutes a genuine two-sector extension of  
holographic dark energy, with known scenarios embedded as consistent subcases.

The framework developed here opens several directions for future investigation.
At the phenomenological level, a full statistical analysis using observational  
datasets would allow quantitative constraints on the entropic exponents  
and a comparison with $\Lambda$CDM and other holographic models.  
At the theoretical level, it would be interesting to explore the implications  
of the two-parameter entropy for early-Universe dynamics, black-hole  
thermodynamics, and the gravity-thermodynamics correspondence more generally.  
Finally, since the entropic functional originates from a microscopic  
generalization of statistical mechanics, further study of its underlying  
information-theoretic and quantum-gravitational interpretation may shed  
additional light on the holographic origin of cosmic acceleration.

\section*{Acknowledgements} 
The research of GGL is supported by the postdoctoral funding program of the 
University of Lleida. The authors  acknowledge  the contribution of the LISA 
CosWG, and of COST Actions CA21136 ``Addressing observational tensions in 
cosmology with systematics and fundamental physics (CosmoVerse)'', CA21106 
``COSMIC WISPers in the Dark Universe: Theory, astrophysics and experiments'', 
 CA23130 ``Bridging high and low energies in search of quantum gravity 
(BridgeQG)'', and CA24101
 ``Testing Fundamental Physics with Seismology''.

 \bibliographystyle{apsrev4-1}
\bibliography{Bib2}

\end{document}